\documentclass{article}

\usepackage{arxiv}

\usepackage[utf8]{inputenc} % allow utf-8 input
\usepackage[T1]{fontenc}    % use 8-bit T1 fonts
\usepackage{hyperref}       % hyperlinks
\usepackage{url}            % simple URL typesetting
\usepackage{booktabs}       % professional-quality tables
\usepackage{amsfonts}       % blackboard math symbols
\usepackage{nicefrac}       % compact symbols for 1/2, etc.
\usepackage{microtype}      % microtypography
\usepackage{lipsum}
\usepackage{graphicx}
\graphicspath{ {./images/} }

\title{Layout2Rendering: AI-aided Greenspace design}

\author{
 Ran Chen \\
  School of Landscape Architecture\\
  Beijing Forestry University \\
  Beijing, 100083 China \\
  \texttt{chenran705367787@bjfu.edu.cn} \\
   \And
 Zeke Lian \\
  School of Landscape Architecture\\
  Beijing Forestry University \\
  Beijing, 100083 China \\
  \texttt{lianzeke@nbcc.edu.cn} \\
  \And
 Yueheng He \\
  School of Landscape Architecture\\
  Beijing Forestry University \\
  Beijing, 100083 China \\
  \texttt{Candyhyh808@bjfu.edu.cn} \\
  \And
 Xiao Ling \\
  School of Landscape Architecture\\
  Beijing Forestry University \\
  Beijing, 100083 China \\
  \texttt{lingxiao@bjfu.edu.cn} \\
  \And
 Fuyu Yang \\
  School of Landscape Architecture\\
  Beijing Forestry University \\
  Beijing, 100083 China \\
  \texttt{yangfuyu@bjfu.edu.cn} \\
  \And
 Xueqi Yao \\
  School of Landscape Architecture\\
  Beijing Forestry University \\
  Beijing, 100083 China \\
  \texttt{Yxq18210187099@bjfu.edu.cn} \\
  \And
 Xingjian Yi \\
  School of Landscape Architecture\\
  Beijing Forestry University \\
  Beijing, 100083 China \\
  \texttt{yixingjian0906@bjfu.edu.cn} \\
  \And
 Jing Zhao\footnote{\texttt{zhaojing@bjfu.edu.cn}} \\
  School of Landscape Architecture\\
  Beijing Forestry University \\
  Beijing, 100083 China \\
  \texttt{zhaojing@bjfu.edu.cn} \\
}
\begin{document}
\maketitle

\begin{abstract}
In traditional human living environment landscape design, the establishment of three-dimensional models is an essential step for designers to intuitively present the spatial relationships of design elements, as well as a foundation for conducting landscape analysis on the site. Rapidly and effectively generating beautiful and realistic landscape spaces is a significant challenge faced by designers. Although generative design has been widely applied in related fields, they mostly generate three-dimensional models through the restriction of indicator parameters. However, the elements of landscape design are complex and have unique requirements, making it difficult to generate designs from the perspective of indicator limitations. To address these issues, this study proposes a park space generative design system based on deep learning technology. This system generates design plans based on the topological relationships of landscape elements, then vectorizes the plan element information, and uses Grasshopper to generate three-dimensional models while synchronously fine-tuning parameters, rapidly completing the entire process from basic site conditions to model effect analysis. Experimental results show that: (1) the system, with the aid of AI-assisted technology, can rapidly generate space green space schemes that meet the designer's perspective based on site conditions; (2) this study has vectorized and three-dimensionalized various types of landscape design elements based on semantic information; (3) the analysis and visualization module constructed in this study can perform landscape analysis on the generated three-dimensional models and produce node effect diagrams, allowing users to modify the design in real time based on the effects, thus enhancing the system's interactivity. 
\end{abstract}

% keywords can be removed

\section{Introduction}
Landscape planning and design are deemed crucial tasks, contributing to ecological equilibrium maintenance, and enhancing people's quality of life. In the conventional landscape design process, crafting a three-dimensional model of the site stands as a pivotal stride for designers. This endeavor not only elucidates the spatial interrelations of design components, facilitating designers in intuitively grasping the park's scale, but also furnishes a groundwork for ensuing landscape analysis and visualization stages. Nonetheless, in practical applications, the entire trajectory, spanning from plan conception to drafting, modeling, and finally rendering, exacts a considerable time toll on designers. Hence, an automated generative 3D park methodology emerges as a swift conduit to augment the efficacy of human designers.

In recent years, propelled by advancements in AI technology, numerous scholars have delved into generative design. Among these, many resort to evolutionary algorithms and parameterization techniques to engender models. A plethora of generative design studies endeavors to fine-tune indicator parameters to steer the generation of 3D models. Nonetheless, this approach encumbers landscape design, and presently, none have applied it to the realm of landscape architecture for three-dimensional model generation. Within the customary design workflow, owing to the intricate confluence of factors inherent in landscape design and its bespoke exigencies, designers frequently draw inspiration from analogous completed designs rather than constraining themselves within the confines of indicator-based design. In architectural circles, researchers have proffered a procedural framework, spanning from plan layout generation through information vectorization to model block instantiation, to address this quandary. Nevertheless, its current applicability is confined to rudimentary architectural block models, rendering the diverse elements of landscape design a daunting challenge. Addressing these predicaments, this study proffers a generative park space design process. Postulating the plan generation design predicated on the topological constraints amid geometric elements, models are subsequently engendered grounded in vectorized information, expeditiously traversing a gamut from fundamental site conditions to model effect fruition.

The principal contributions of this study encapsulate: (1) Propounding a swift workflow tailored for landscape design, where users input site conditions to expeditiously generate design models; (2) Evidencing the expeditious generation of spatial green space schemes aligned with the designer's perspective leveraging AI-assisted technology, with concomitant model analysis and real-time modifications under human-computer interaction; (3) Pioneering a model analysis and visualization system to furnish a beacon for the analysis and evaluation of computer-generated outcomes within the landscape architecture domain; (4) Proffering a methodology for extracting and Three-Dimensional disparate landscape design element vector structures while retaining design fidelity.

The ensuing sections of this paper are organized as follows: Section 2 furnishes a comprehensive literature review of related endeavors. Section 3 delineates our framework and experimental modalities. Section 4 expounds upon the findings of the experiments. Section 5 engenders a discourse on pertinent research. Section 6 proffers conclusions and outlines future avenues of inquiry.

\section{Related work}
\label{sec:headings}
Our related work encompasses vector edge extraction and generative design. In this section, we primarily review the pertinent literature in these two areas.

\subsection{Vector edge extraction}
In the field of computer vision, image storage can generally be classified into two main forms: bitmap and vector graphics. Bitmap relies on pixel arrays to construct images, with its main drawback being the tendency to become blurry and lose detail clarity upon enlargement. In contrast, vector graphics depict shapes by defining the length and direction of lines and curves, offering flexibility in editing and maintaining image clarity and detail integrity regardless of scaling. Moreover, owing to its exceptional software compatibility, vector graphics have found widespread application in the design domain, serving as a crucial medium for research data flow dynamics.

Image vectorization, the conversion from bitmap to vector graphics, encompasses two subprocesses: special type image vectorization \cite{zhang2009vectorizing} and natural image vectorization \cite{inproceedings}. Special type images, such as sketches, clip art, cartoons, character images, and engineering drawings, primarily focus on extracting element boundaries and optimizing them through parameterized curves due to their simple color and structure.

Current research in image edge detection methods can be divided into traditional methods and deep learning-based methods. Traditional methods primarily rely on image gradients or second-order derivatives to detect edges, usually enhancing the extraction capability for image textures and colors through manual feature extraction and edge classifier training. At this stage of research, the commonly used edge extraction methods are based on threshold segmentation, based on edge monitoring segmentation and so on. The methods of image edge detection can be divided into traditional methods and deep learning based methods. Traditional methods are mainly based on the gradient or second-order derivatives of the image to detect edges, and the common operators are Roberts, Prewitt, Kirsch, Laplace, Sobel \cite{tiis:23271}, Canny \cite{sym12111749} and so on. These methods are simple and easy to implement, but there are some drawbacks, such as sensitivity to noise, discontinuous edges, and inaccurate localisation. In order to overcome these drawbacks, some improved methods have been proposed.Wu et al. \cite{article} proposed a colour image edge extraction method called function gradient, which segments the target image edges based on colour features. While in architectural edge extraction, Liu et al. \cite{liu2022lcs} proposed the framework of Line Segment Collaborative Segmentation (LCS) to provide positional guidance for vector extraction to obtain the exact position of the boundary. In 2003, Daniel, Sykor et al. \cite{10.1145/984952.984989} employed the Laplacian of Gaussian (LOG) operator to extract edge features of cartoon images but failed to effectively distinguish decorative lines and edge features, and was limited to recognizing black and white cartoon images. In 2011, Yang et al.  \cite{ping2011unbiased} addressed the problem of the Sykora algorithm's inability to detect decorative lines by employing Delaunay triangulation, thus improving contour recognition accuracy. In 2006, Swaminarayan et al. \cite{swaminarayan2006rapid} developed the RaveGrid vectorization system, which processed original raster images using methods such as image segmentation, edge extraction, color sampling, and triangulation to generate vector images. The ARDECO vectorization system developed by Lecot et al. \cite{10.2312:EGWR:EGSR06:349-360} replaced the method of triangulation with cubic spline fitting to process images, and regions enclosed by fitted curves were filled with colors for rendering vector images. In 2011, Ravi et al. \cite{0c1404cfeb114ed3862d9f1d02ae31a5} further optimized the vectorization recognition algorithm, which preserved the sharp features of the original image while better expressing geometric elements such as curves and straight lines, making the vectorized images more stretchable and compressible. Concurrently with the development of image vectorization technology, this field began to focus on the impact of resolution on vectorization recognition results, especially low-resolution image recognition, which has always been a challenging problem in this field. In 2011, Kopf et al. \cite{Kopf2011} proposed a solution to the low-resolution image vectorization problem by optimizing the fitting of identified contour curves, using piecewise smooth curves to fit recognized contours, which improved the smoothness of edge contour lines and enhanced the visual effect of vector images.

In recent years, with the outstanding performance of convolutional neural networks in extracting image features, deep learning methods have made significant advancements in the field of image processing. This approach has addressed many issues of traditional methods, such as continuity and noise resistance, while simplifying the detection process without the need for manual feature design. For example, inspired by SE, Lempitsky \cite{ganin2014} proposed the N4-Fields edge detection algorithm based on convolutional neural networks (CNNs). This algorithm combines convolutional neural networks and nearest neighbor search (NNS) to classify image block features, thereby obtaining similar contours. Additionally, the DeepContour algorithm achieved learning of edge shapes by transforming the binary classification problem into a multi-class problem \cite{Shen_2015_CVPR}. Among the end-to-end neural network-based methods, the HED model combined multi-scale learning with rich hierarchical features and was trained and predicted in an image-to-image manner \cite{Xie_2015_ICCV}. In 2022, Pu et al. \cite{Pu_2022_CVPR} first introduced the attention mechanism (Transformer) into edge detection and proposed the ENTER model, which outperformed various detection methods in the BSDS500 edge detection dataset test. Transformers have been widely used in the field of vision and demonstrate better performance in edge detection across various domains. Edge detection models based on Transformer's self-attention mechanism have become a hot research direction for the future.

In summary, traditional edge detection methods have matured in aspects such as image edge recognition and edge line fitting optimization and have wide applications in multiple fields. However, there is a lack of targeted vector edge extraction methods in the field of park green space plan design. Although deep learning-based edge detection methods offer excellent extraction results, they rely on many high-quality datasets for training. Moreover, due to the complexity and interpretability issues of neural network structures, they present some challenges and limitations.

\subsection{Generative design}
Generative Design is a design methodology based on computer algorithms and rule systems that has been widely applied in the field of design with the advancement of technology. Initially, it relied on parametric techniques, where designers provided rules to achieve optimal solutions in forms of "self-organization" and "self-optimization" for simple design objectives. As evolutionary algorithms evolved, cellular automata, genetic algorithms, shape grammars, multi-agent systems, among others, gradually emerged to tackle complex design problems. In recent years, the rapid development of deep learning has ushered in a new phase of development for generative design, with Generative Adversarial Networks (GANs) emerging as a key driver of progress in the field of artificial intelligence.

Generative Adversarial Networks (GANs) are a type of neural network framework proposed by Goodfellow et al. in 2014 \cite{NIPS2014_5ca3e9b1}. Its uniqueness lies in comprising two competing components: the generator and the discriminator. The generator's task is to produce realistic-looking virtual data from random noise, while the discriminator's task is to distinguish whether the input data is real or fabricated. During the training process, these two components adversarial compete, striving to reach a balance where the generator can generate virtual data that deceives the discriminator.

The GAN-based design method was initially applied to the generation of architectural interior space layouts. Huang and Zheng \cite{11596c081f5c40f899f1bb97cda45183} used pix2pix to identify and generate architectural drawings, exploring the layout of interior space plans, thus establishing the main research paradigm in this field. Subsequently, this application method gradually entered the fields of architectural design and urban planning, mainly used for the automatic generation of building and urban layouts. In outdoor space design, the spatial complexity and flexibility of park green spaces are higher. Currently, the research paradigm for generating park green spaces design is generally limited to "site layout planning." For instance, Liu et al.  \cite{inbook} adjusted the layout label set to guide the generation of layouts for Jiangnan private gardens. Chen et al. \cite{buildings13041083} implemented the generation of park green space layouts and the rendering of color plans based on pix2pix and CycleGAN[9]. In the field of architecture, the detailed information stored in 3D models is generally complex, making it difficult to directly convert it into input data for Generative Adversarial Networks to output 3D models. In architectural design, researchers have proposed a process framework from plan layout generation to information vectorization to model block establishment to address this issue, but it can only be applied to generating basic architectural block models. In the field of park green space design, due to the complexity of element types, irregularity of most element edges, and complexity of elevation variations in park green space 3D models, there is still a technical gap, and a systematic process framework for generating park green space 3D models has not been achieved.

Currently, generative design can be implemented through computer algorithms and manually devised rule systems; in recent years, Generative Adversarial Networks (GANs) in deep learning have been widely applied in this field, showing potential, especially in architecture design and urban planning. However, in the field of park green space design, generative design is still limited to plan layouts, and there is a technical gap in 3D model generation.

\section{Methodology}
\subsection{Analytical framework}
\begin{figure}
    \centering
    \includegraphics[width=\textwidth]{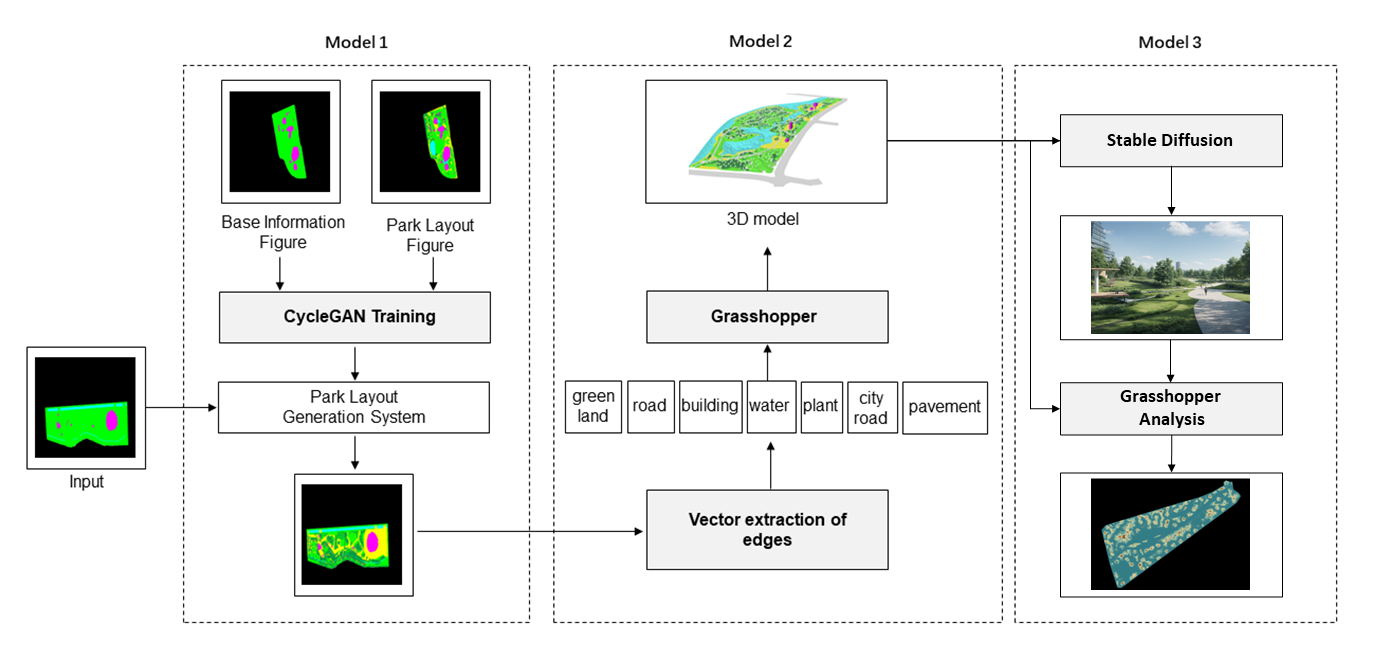} % 将图像宽度设置为文本宽度的80%
    \caption{Overall analytical framework.}
    \label{fig:fig1}
\end{figure}

As depicted in Figure \ref{fig:fig1}
, the proposed generative park design framework in this paper comprises three phases. The content of each phase is outlined as follows:

The first phase involves plan generation design. An automated plan generation system is developed based on deep learning technology, capable of rapidly generating a large number of complete design layout drawings solely constrained by the layout of internal park elements.

The second phase is 3D park automated generation design. Building upon the data generated by the plan generation system in the first phase, a parametric modeling system is developed to swiftly construct three-dimensional models.

The third phase focuses on analysis and visualization. A model processing system is constructed to enable rapid analysis and modification of the generated three-dimensional parks, producing visual renderings from different perspectives. Subsequent sections will delve into each step in detail.

\subsection{Plan layout generative design based on GAN}
In this phase, we adopted the framework proposed by Chen et al.  \cite{buildings13041083} and utilized deep learning technology to train a plan generation model using park design images and corresponding layout labels. We obtained 194 pairs of park plan design images from the internet and annotated their elements with different colors. The element categories were divided into eight classes: green spaces, water, roads, pavements, buildings, red lines, city roads, and plants. These images were then processed into 256*256 pixel sizes for training. Following Chen's research, we employed the CycleGAN model to augment the 194 element layout images to 4047, thus establishing an automated plan generation design system.

\begin{figure}
    \centering
    \includegraphics[width=\textwidth]{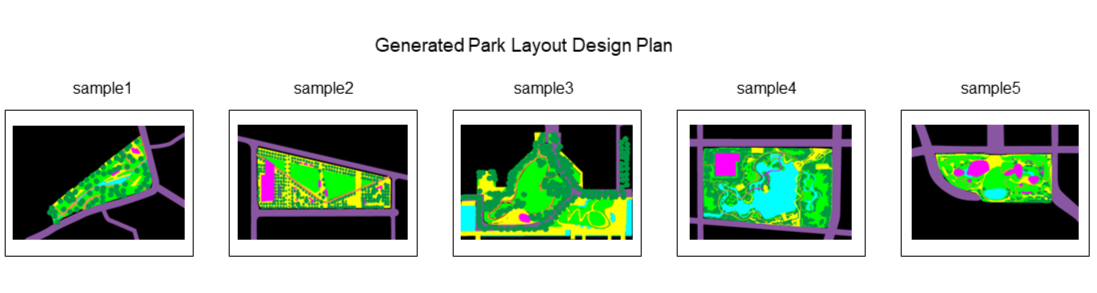} % 将图像宽度设置为文本宽度的80%
    \caption{Generated park layout design plan.}
    \label{fig:fig2}
\end{figure}

Based on the trained landscape plan generation model, we input numerous layout images with given constraints and obtained corresponding layout plan design images. From these, we selected the five most authentic and distinctive design schemes as subsequent experimental data samples, as shown in Figure \ref{fig:fig2}. The selection criteria for the samples were as follows: (1) encompassing various scales of parks; (2) featuring diverse vegetation planting methods; (3) exhibiting a variety of arrangements of elements such as pavements and water.

\subsection{Data processing}
Data processing in this section consists of two parts: image preprocessing and vector edge extraction.

\subsubsection{Image preprocessing}
Due to the uneven color distribution within color blocks and the presence of jagged edges along block boundaries in layout images directly generated by GANs, these issues can interfere with the segmentation and extraction of elements. Prior to extracting edge features from images, it is necessary to minimize the influence of noise within the images. Utilizing the Lo gradient minimization model, noise within the images is reduced or blurred through image smoothing techniques. Subsequently, image enhancement techniques are employed to sharpen edge features, thus enhancing the accuracy and efficiency of image processing tasks.

In the task of image smoothing, we manipulate pixel values within the image through averaging, filtering, and other processes while retaining essential features to eliminate noise. Essentially, algorithms are employed to adjust pixel values that differ significantly from surrounding pixels to approximate values, reducing unnecessary jaggedness and sharp angles in contours, thereby improving the accuracy and effectiveness of vectorization.

On the other hand, image enhancement aims to improve image quality and recognizability. To better display useful edge information within images, sharpening techniques and contrast enhancement are applied to highlight edge features and enhance clarity.

\subsubsection{Vector edge extraction}
Vector edge extraction is an image processing technique used to detect clear edges in images. In this section, we first conducted image segmentation experiments based on the color characteristics of plan elements, followed by contour vectorization recognition of different design elements.

In two-dimensional generative design, we can experimentally identify image features from raster data. However, in three-dimensional generative research, converting two-dimensional elements into three dimensions requires certain vector information. Unlike traditional pixel-level edge segmentation, image vector edge extraction focuses on generating vector information that expresses edge direction and position. Since the segmentation task of this study aims to extract design elements from plan images, and in our data samples, different design elements correspond to different RGB values, we chose the OpenCV threshold segmentation method for simple and efficient extraction of image information.

The OpenCV segmentation method based on color characteristics is as follows: First, we set up the Python environment library and import the OpenCV toolkit. Then, we convert the image from BGR format to RGB format. Next, we create an array image() with the same height and width as the input image and fill it with a matrix of all 0s for RGB values. The algorithm then iterates through each color range defined by each design element category, scanning each pixel of the input image to determine if it falls within the defined range. Pixels within the same range will be displayed in the matrix with RGB values of 255 and saved as a PNG black-and-white segmentation image.

\begin{figure}
    \centering
    \includegraphics[width=\textwidth]{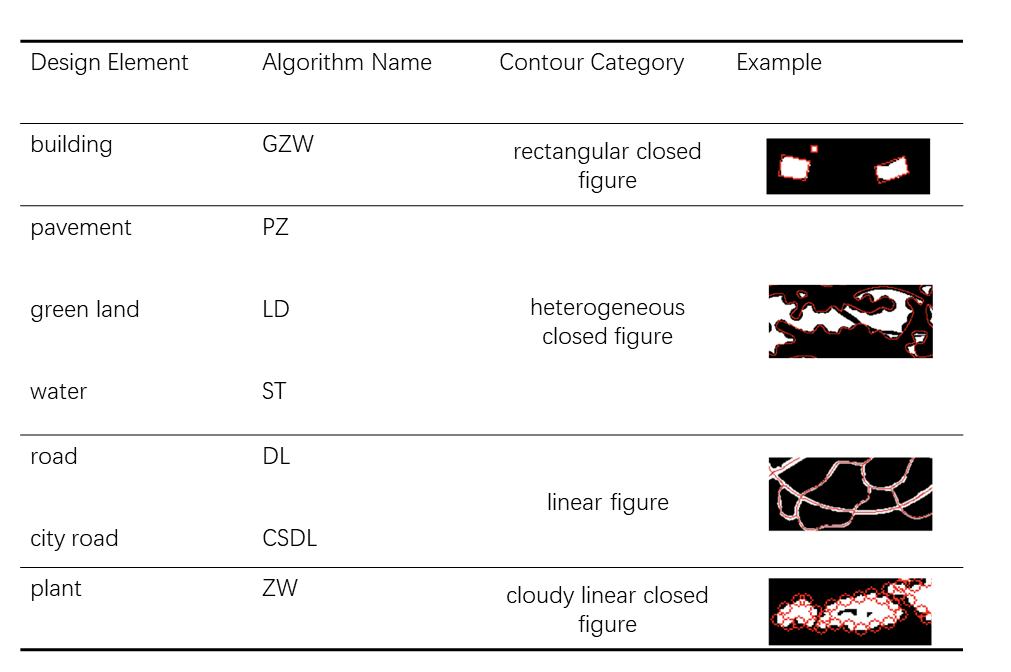} % 将图像宽度设置为文本宽度的80%
    \caption{Categorisation of design element profile features.}
    \label{fig:fig3}
\end{figure}

\begin{figure}
    \centering
    \includegraphics[width=\textwidth]{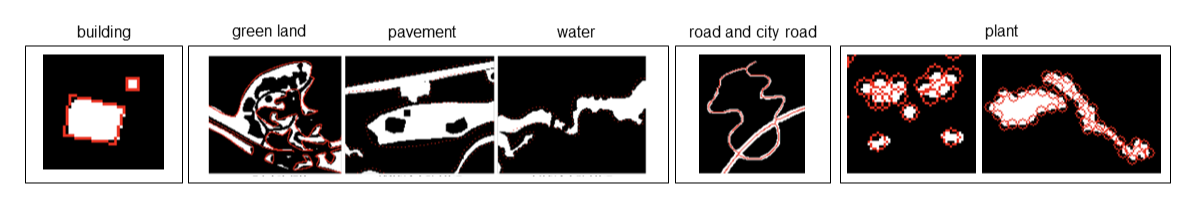} % 将图像宽度设置为文本宽度的80%
    \caption{Optimisation effect diagrams for different elements.}
    \label{fig:fig4}
\end{figure}

After segmenting each category of elements, we optimize the algorithm based on the different characteristics of design elements, recognizing contours and fitting curves to obtain vector data from the segmentation results. However, the elements from which we need to extract edges are design elements in the field of landscape architecture, which not only contain image contour information but also crucial landscape design features and information. Therefore, different processing methods are needed for different elements. We categorize the seven types of graphical elements that need to be extracted into four different categories based on their characteristics, as shown in Figure \ref{fig:fig3}: (1) Rectangular closed figures: building;(2) Heterogeneous closed figures: pavement, green land, water; (3) Linear figures: roads, city roads; (4) Cloudy linear closed figures: plants. Figure \ref{fig:fig4} shows the post-processing results for each element.

\paragraph{(1)Algorithm optimization for rectangular closed figure:}
In landscape design, structures such as pavilions, bridges, towers, walls, and corridors often adopt linear structures such as vertical and parallel lines. However, such buildings may be lost due to low resolution in image segmentation. Therefore, we employ rectangle edge fitting to restore the original structural features of the buildings. Since the buildings in the layout may not be aligned along the XY axis, and their outer edges may not be regular rectangles, the algorithm first checks if the contours have been rotated. If so, it calculates the minimum bounding rectangle of the contour and redraws the vector graphics based on the four vertices. If not, it directly calculates the bounding box of the parallel coordinate axis. Finally, the optimized vector edges are saved to a JSON file.
\paragraph{(2)Algorithm optimization for irregular closed figure:}
Elements such as green land, pavement, and water in landscape design are characterized by their free-flowing lines and dynamism. However, these characteristics may be difficult to express in image segmentation due to the jaggedness of the edges, requiring noise elimination and smooth fitting of the edges by the algorithm. We propose an edge smoothing optimization algorithm that utilizes convolution kernels for dilation and erosion to eliminate noise. Subsequently, the outer contours are extracted, and contours smaller than a specific area are filtered out. For the retained contours, we use polygon fitting and Bezier curve smoothing, followed by sparsification of the fitting points to significantly reduce the number of points. Finally, the processed contours are drawn back onto the original image, and the contour coordinate information is saved.
\paragraph{(3)Algorithm optimization for linear figures:}
For elements such as roads that require maintaining shape information and connectivity, we propose a new optimization algorithm that preserves the most important feature of the road, the road centerline. First, the low-resolution image is subjected to Gaussian blur to eliminate noise and then converted to grayscale for binarization. Next, the thinning function is applied to extract the skeleton of the image to highlight the road centerline feature. Contours are extracted from the thinned image using the findContours function, and contours smaller than a specific size are filtered out. Filtering and sparsification further refine the contours. Finally, the extracted contour lines are drawn back onto the original image, and the contour point coordinates are saved.
\paragraph{(4)Algorithm optimization for cloudy linear closed figures:}
Planting designs in the field of landscape architecture often come in various forms to consider the coordination between plants and the environment for overall aesthetics. Typically, a single tree in the plan is represented as a circle with the tree trunk position as the center and the average radius of the tree crown as the radius. The representation mainly consists of multiple separate circular tree circles and cloudy linear plantings. However, segmentation algorithms can only identify vegetation contours and cannot recognize the complete outer edges of individual tree clusters and planting points, significantly affecting the reproduction of the real design process and causing interference with subsequent modeling and analysis processes. Our optimization algorithm introduces Hough circle detection to change the output result from contour line fitting point information to planting information. First, contours are extracted, and for contours of different areas, the algorithm processes single plantings (contours smaller than 30 pixels) and cloudy linear contours (contours greater than or equal to 30 pixels) separately. For single plantings, the minimum enclosing circle is determined, and its center coordinates are recorded. For cloudy linear contours, planting points are extracted at intervals to enhance the realism and detail representation of the design. Finally, the algorithm saves and outputs the processed images and their contour coordinate information.

\subsection{Parameter-based 3D model generation}
In this section, the study creates corresponding 3D models based on the extracted vector data to explore the potential of vectorized data acting as a "data stream." The entire process is divided into two steps: vector database construction and automatic modeling system construction.

\subsubsection{Vector database construction}
After extracting vectorized data for different elements in the previous stage, we utilize the Python interface in the Grasshopper platform to read contour point information and organize the vectorized information of landscape elements into different information interfaces, constructing a complete database for the subsequent tasks. Below are the specific construction methods.

First, import the RhinoScriptSyntax, os, and json modules in the Python platform of Grasshopper. Build a data tree structure to store and manipulate data. After reading the vectorized information of design elements, iterate through the 7 categories of design element files, extract the data points, and convert them to Rhino3D format. Finally, store the converted data points in the DataTree, which will be stored in different variables such as CSDL, DL, and GZW for subsequent use.

\subsubsection{Construction of automated modeling system}
\begin{figure}
    \centering
    \includegraphics[width=\textwidth]{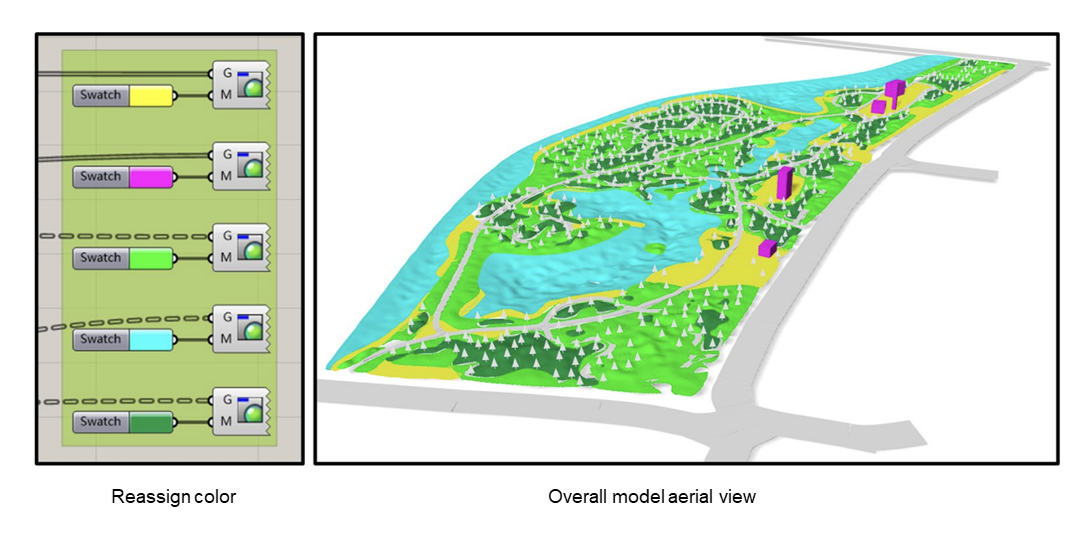} % 将图像宽度设置为文本宽度的80%
    \caption{Overall model coloring battery pack with aerial view.}
    \label{fig:fig5}
\end{figure}

This study utilizes the Grasshopper platform to parametrically generate models based on processed vectorized data. In the fields of architecture and planning, generative design mainly involves the construction of volumes on different spatial vectors based on the constraints of objective indicators. However, due to the complexity of landscape architecture design, it is difficult to generate designs through indicator control. Additionally, different elements contain spatial information in different ways. Therefore, we have developed separate 3D modeling methods for the design features of each element. The generated model effects are shown in Figure \ref{fig:fig5}.

\paragraph{(1)Buildings:}
In landscape design models, the most important aspect of buildings is to represent spatial relationships. Therefore, this study adopts a simplified approach to represent buildings using concise polygonal blocks. Firstly, the coordinate point information of buildings is imported from the database. Then, after constructing the base with polylines and boundary surfaces, random values are assigned within the regular building height range, and extrusion is applied to obtain simplified buildings.
\paragraph{(2)Pavement:}
In the 3D model, pavement is also primarily used to represent positional contours and their relationships with the surroundings. Using the same operations as above, after importing the pavement coordinate information, duplicate overlapping points are removed, and pavement surfaces are constructed.
\paragraph{(3)Water and Green land: }
As important elements in the 3D landscape model containing elevation information, the undulating terrain and free-form contours of water and green land create a unique spatial experience in landscape environments. However, current AI models can only recognize planar elements and generate two-dimensional data. Therefore, in this study, the battery assembly function relationship is utilized to represent the elevation changes in three-dimensional space separately for water and green land. After establishing surfaces for water and green land using the same method as pavement, they are processed into triangular mesh structures. Elevation information is established by moving grid points, with water data moving negatively along the z-axis and green land moving in the opposite direction. Since lakes and green land in reality typically have forms that gradually change in height from the edge to the center, reasonable elevations are randomly assigned while coupling the growth of shoreline distances. Finally, the moved points are converted into a three-dimensional network model together with the initially generated surfaces.
\paragraph{(4)Roads:}
Since roads primarily represent the connection between various spaces in landscape design, and different types of roads usually have certain standard widths, this study directly extracts the centerlines of roads from the imported data and offsets them to generate roads of equal width. Meanwhile, different widths are assigned to city roads and garden paths.
\paragraph{(5)Plantings:}
In the vectorization extraction step, plant layouts were divided into single and clustered groups. In the model generation process, we follow the same grouping. For single plantings, after importing the center point coordinate data, tree models are directly generated at planting points using the tree model function in the Lands Design for Rhino plugin. For clustered plantings, since vectorized data can only fit the outer contours of planting areas and cannot obtain the center points of each tree, this study estimates the tree center points based on the ratio relationship between the number of planting points and the planting area. We use the Division command to control planting density, ensuring a certain coupling relationship between the number of plantings and the planting area. This completes the generation of all tree models.

\subsection{Analysis and visualization}
In most existing generative design studies, qualitative indicators are often used to evaluate the results of generative models. For instance, Yoo et al. \cite{Yoo} evaluated the engineering performance of generated 3D wheels under different stiffness and frequency conditions. In urban planning, Gao et al. calculated the daylighting capacity of buildings based on parameters such as building height and plot size to find optimal balance solutions for addressing daylighting issues in high-density cities. However, in the field of landscape planning, generative design poses challenges due to the complexity of spatial relationships among landscape design elements, making it difficult to assess directly using simple quantitative indicators. Typically, in actual design processes, designers continuously modify designs through multiple interactions with clients until satisfactory results are achieved. Given this challenge, we have constructed an analysis and visualization system to intuitively demonstrate the design's effects and its impact on the urban environment, facilitating rapid understanding and modification of designs by various stakeholders.

In this stage, we first evaluate the most fundamental step in landscape design—"elevation"—to visually reflect the terrain conditions of the generated model excluding water, assisting designers in understanding the approximate terrain conditions of the site and exploring possibilities for terrain design. In this step, we utilize Grasshopper for visual analysis of elevation-related aspects of the generated model, including elevation, slope, and drainage.

\begin{figure}
    \centering
    \includegraphics[width=\textwidth]{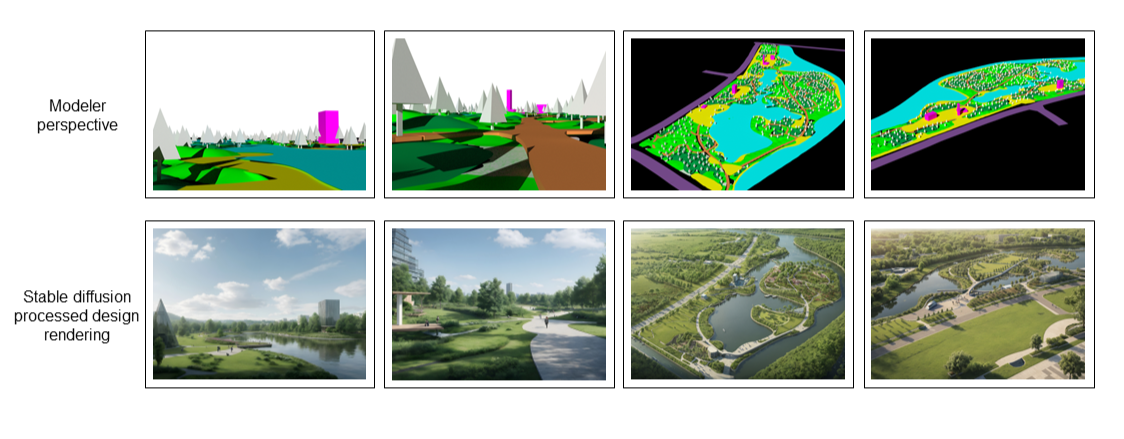} % 将图像宽度设置为文本宽度的80%
    \caption{Stable Diffusion's results for different viewpoints.}
    \label{fig:fig6}
\end{figure}

\paragraph{(1)Elevation analysis:}
After extracting all Z-vector information of the generated model's terrain surfaces, we conduct internal comparative analysis of elevation visual surfaces reflecting color changes with elevation variations. By analyzing the distribution of color information in relation to the distribution of roads, pavement, and water, we assess the reasonableness of the elevation and provide information for designers to explore site possibilities and modify designs.
\paragraph{(2)Slope analysis:}
We segment the formed terrain, extract the normal vectors of the cutting surfaces, and compare them with the Z-axis vectors. We use five slope values—0 degrees, 5 degrees, 19 degrees, 45 degrees, and 65 degrees—related to human perception as boundary values for four filters. Surfaces that meet the conditions are selected and filled with different colors to visualize the slope of the generated model, aiding designers in modifying and exploring spatial elevations of the site.
\paragraph{(3)Drainage analysis:}
Finally, we conduct site drainage analysis using the Mosquito plugin on the generated model, reflecting the drainage situation of the generated model. This allows designers to intuitively and conveniently adjust the site elevation based on the distribution analysis of water, roads, pavement, and buildings to improve the effectiveness of rainwater drainage.
Subsequently, to provide designers with more references during the modification phase, we select multiple human perspectives and bird's-eye views within the generated model and combine them with Stable Diffusion to generate bird's-eye views and node effect diagrams of the design, offering more insights. Some of the design diagrams generated by Stable Diffusion are shown in Figure \ref{fig:fig6}.

\section{Experimental results}
In this experiment, we trained a CycleGAN model for generating park layouts and selected five sets of samples to compare the results of model generation, aiming to evaluate the universality of 3D model generation. The criteria for sample selection were as follows: (1) images covering parks of different scales; (2) images including water and green spaces of different shapes and sizes; (3) images featuring diverse planting methods; (4) images with buildings arranged in different layouts.

This experiment completed 8 modeling logic paths for 7 types of design elements. The final complete model was largely consistent with the layout plan, and after the Bake command, it could be used for multiple purposes. Each design element was recolored according to the RGB channels, allowing different areas to serve as base maps for renderings. Suitable perspectives were selected to process and refine the model using Stable Diffusion and Grasshopper. On one hand, the model itself could be directly imported into rendering software such as Lumion and Enscape for rendering, and the model accuracy could meet the requirements for expressing design information from bird's-eye views. On the other hand, a portion of the human perspective could be selected and rendered to obtain renderings suitable for project use.

\begin{figure}
    \centering
    \includegraphics[width=\textwidth]{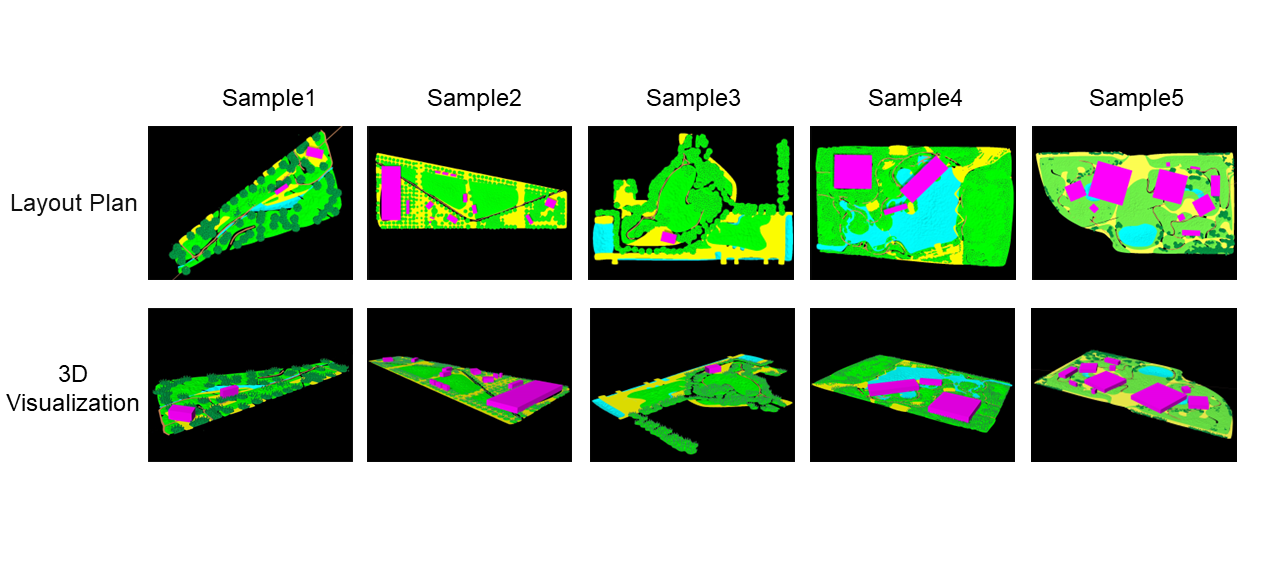} % 将图像宽度设置为文本宽度的80%
    \caption{The results of the plan layout and its 3D model generation.}
    \label{fig:fig7}
\end{figure}

In the experimental results of model generation, all five samples exhibited good generation effects, as shown in Figure \ref{fig:fig7}
. It can be observed that this parameterized generation method could generate roads and buildings of reasonable scales in parks of different scales (e.g., Sample1). For the generation of water and green spaces of different shapes and sizes, it could present a sense of spatial enclosure and terrain coherence (e.g., Sample4,5). However, there were also certain shortcomings: (1) For tree generation, the model's forms and species were relatively monotonous, and it could not handle the edge lines of forests from a design perspective. (2) Some roads and pavements could not adapt to terrain changes, resulting in model overlaps (e.g., Sample1,3).

\begin{figure}
    \centering
    \includegraphics[width=\textwidth]{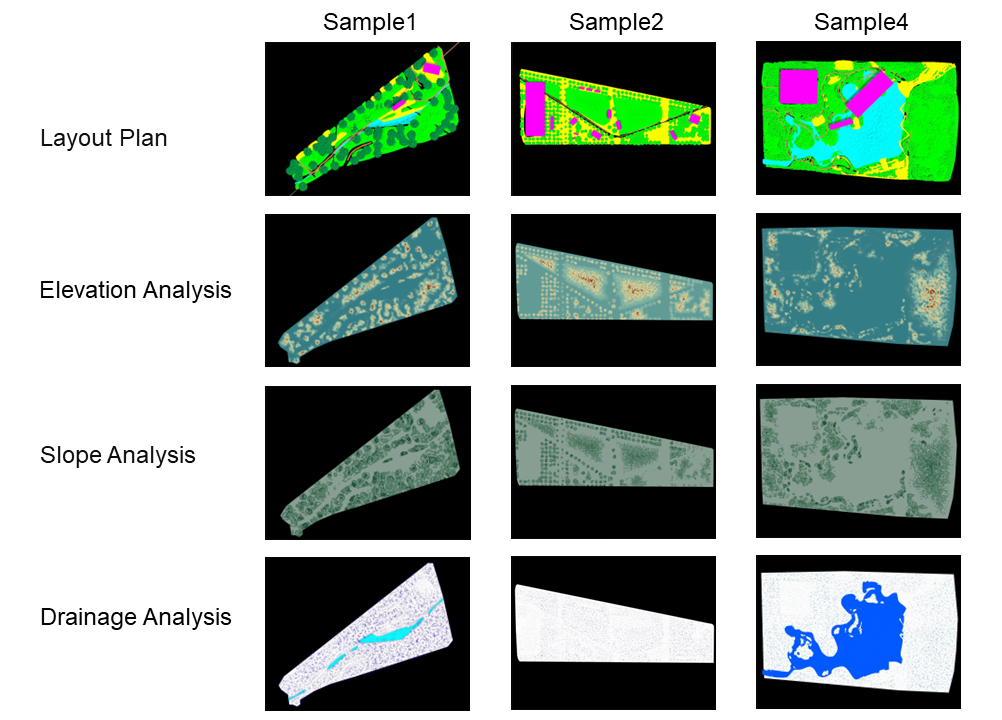} % 将图像宽度设置为文本宽度的80%
    \caption{The results of the plan layout and its 3D model generation.}
    \label{fig:fig8}
\end{figure}

From the test results of model analysis, as shown in Figure \ref{fig:fig8}, we can see that the elevations of all models were within a realistic range. The elevation analysis of parks revealed the variations in natural terrain features within the park, such as green spaces and water systems. The results can help planners determine the optimal locations for viewpoints, rest areas, and landscape features to fully utilize natural terrain and enhance visitor experiences. For example, observation decks or pavilions can be set up at high points to provide panoramic views of the park and surrounding areas. Slope analysis of the park can visually display the degree and direction of ground slope in different colors, which is crucial for planning the locations of landscape features. Additionally, slope information is important for preventing soil erosion and guiding vegetation planting and land management strategies. Moreover, drainage analysis helps determine the paths of rainwater and surface water flow, aiding designers in effectively constructing park drainage systems and laying the foundation for creating rain gardens.

\section{Discussion}
The paper presents an innovative park space generation system that integrates generative adversarial networks (GANs) with Grasshopper, effectively achieving the rapid conversion of site plan basic information into three-dimensional models. This system further supports basic landscape analysis and the generation of node effect diagrams, constructing a complete intelligent design process. This not only enhances the interaction between designers and models but also promotes the progress of human-machine collaboration in the landscape industry.

In the field of planning and design, generative design is rapidly evolving. The most common method is parametric modeling based on Grasshopper, where researchers often use urban generation as an example, introducing key performance indicators such as energy efficiency and indoor daylighting control. For instance, Julian et al. \cite{ZARAZA2022104274} seek the optimal balance between high-rise residential design and greenhouse gas emissions by maximizing site efficiency and compliance with visibility and building regulations, effectively reducing energy consumption. Huang et al. \cite{HUANG2022109575} proposed a GAN-based environmental performance-driven urban generation design framework, considering constraints such as thermal comfort and pedestrian wind environments. Yang et al.  \cite{land12061167} generated specific block layouts using shape grammar and Grasshopper, demonstrating a generative urban spatial design approach. These studies emphasize constraints on specific performance indicators, but in landscape design generation, methods that generate by controlling indicators often overlook the unique spatial semantic information between design elements, making it difficult to generate park models that meet actual needs. Additionally, many researchers use methods such as building information modeling (BIM), point cloud generation, and spatial perspective generation for three-dimensional model construction. These methods can provide accurate multi-dimensional information for project construction and maintenance. However, in the landscape architecture field, it is often difficult to generate through these methods due to lack of data.

The system proposed in this study combines the traditional process of landscape design with computational methods, building three-dimensional models of various landscape elements based on the topological relationships between design elements, and promoting the development of the landscape design field from raster data to vector data. The system can not only extract layout rules from numerous actual design cases but also consider the spatial positional relationships of different elements, constructing irregular models based on the semantic characteristics of landscape professionals. Additionally, users can quickly conduct basic terrain analysis and obtain site effect diagrams based on this model.

However, this three-dimensional space generation method has limitations: (1) The generated layout diagram is a raster image, resulting in continuous design elements such as roads and pavements being segmented by plant elements, affecting the accuracy of contour recognition and fitting; (2) The identification of buildings is limited, currently only treating them uniformly as rectangles, ignoring the diversity of building forms in landscape design; (3) In actual landscape design, there are many plant species that require professional horticulturist configuration, which is difficult to consider in current technology for three-dimensional landscape model generation. Although the park space generation system proposed in this study has achieved some results in landscape design automation, further exploration is needed in future work to overcome existing limitations and improve the practicality and accuracy of the system.

\section{Conclusion and future work}
This paper explores the method of vectorization of landscape architecture design outcomes and the circulation of vectorization information in various stages of landscape architecture design practice. The goal is to construct a complete design process covering element layout generation, element vectorization, and three-dimensional modeling, achieving an initial exploration of the vectorization means of landscape architecture design generation and an applicability analysis of vectorized files as data streams in digital landscape architecture.

The main features and innovations of this study include the application of image vectorization technology in the field of landscape architecture design generation, achieving the vectorization identification and generation of design elements. The image vectorization algorithm is optimized and corrected according to the requirements of landscape architecture planning and design, thus outputting vectorization information suitable for three-dimensional modeling and two-dimensional evaluation. This not only greatly expands the application scope of landscape architecture design generation but also provides a new research direction for the vectorization of landscape architecture design element images. Additionally, based on the demand for digital landscape data flow, the vectorization identification of a large number of raster data from landscape architecture design generation enhances the coherence of landscape architecture design stages in the context of digital landscapes. It achieves the circulation of individual design elements of design schemes in multiple design analysis software, effectively improving the efficiency of construction and management processes, reducing the probability of human error, and strengthening the automation and quantification of landscape architecture automatic design.

Future research should further explore means of comprehensive evaluation indicators beyond single area ratio evaluation methods. Meanwhile, alternative evaluation methods for non-planar elements, such as ecological value and road structure, should be explored. Different theoretical approaches should be adopted for the evaluation of different design elements (such as roads, green spaces, water) to ensure that each design element is supported by scientific theory rather than relying solely on prior empirical accumulation.

\bibliographystyle{unsrt}  


\bibliography{template}

\end{document}